\documentstyle[aps,prb,multicol,epsf]{revtex}
\begin{document}
\draft

\input psfig.sty

\title{Crystal-field effects in the first-order valence
transition in YbInCu$_4$ induced by an external magnetic field}
\author{M. O. Dzero}
\address{National High Magnetic Field Laboratory 
and Physics Department, Florida State University, 
Tallahassee, FL, 32304, USA}
\date{\today}
\maketitle
\begin{abstract}
As it was shown earlier {[}Dzero, Gor'kov, and Zvezdin, 
J. Phys.:Condens. Matter {\bf 12}, L711 (2000){]} the 
properties of the first-order valence phase
transition in YbInCu$_4$ in the wide range of magnetic fields and
temperatures are perfectly described in terms of a simple entropy
transition for free Yb ions. Within this approach, 
the crystal field effects have been taken into account and we 
show that the phase diagram in the $B-T$ plane acquires some 
anisotropy with respect to the direction of an external magnetic field. 
\end{abstract}
\begin{multicols}{2}
\narrowtext

As it is well known, YbInCu$_4$ undergoes a first-order valence 
transition at $42$ K, accompanied by a small change in volume
of the order of $0.5\%$. This transition is quite similar to 
the $\gamma-\alpha$ transition in metallic Ce (for the phase diagram
of Ce, see \cite{Koskenmaki}). It turns out, that YbInCu$_4$ is the only
stoichiometric compound known in which an isostructural valence
transition at ambient pressure has been observed \cite{SarraoPC}.
However, as it was pointed out in \cite{SarraoPC}, the isostructural 
valence transition is just the extreme limit of the very common 
competition that occurs between local-moment and itinerant behavior
in many strongly correlated compounds.

The valence transition induced by en external magnetic field 
in YbInCu$_4$ and its alloys has been studied in \cite{Immer1}.
The most interesting result of \cite{Immer1} is that the data
extracted from the resistance measurements can collapse all 
of the pressure dependent data, as well as that from doped variants
of YbInCu$_4$ at ambient pressure, onto a universal $B-T$ phase diagram
(here $B$ is a magnetic field, $T$ is a temperature).

The physics which can be responsible for the transition in 
YbInCu$_4$ has been discussed by several authors. 
One of the first attempts to describe the transition in metallic Ce, 
the one which is similar to the transition
in YbInCu$_4$, was the Falicov-Kimball-Ramirez (FKR) model \cite{Falicov}. 
Another approach in which the $\gamma-\alpha$ transition is ascribed
to the Mott's first-order transition in a subsystem of $f$-electrons
has first been discussed in \cite{Johansson}. 

Very often the ($\gamma-\alpha$) transition in Ce is  
interpreted in terms of the Kondo Volume Collapse (KVC) 
model \cite{Allen,Liu}. In the KVC model Ce atoms at the 
transition are treated as Ce$^{3+}$-ions in the both
$\alpha$ and $\gamma$ phases (approximately one electron 
in the $f$-shell), although in the two different Kondo regimes. 
As it is known, the Anderson impurity model
reproduces the Kondo behavior in the regime when charge fluctuations
are fully suppressed, and provides for the $T_K$ the expression:
\begin{eqnarray}
T_K\propto\exp\left\{-\frac{\mid{\varepsilon_f^{*}}\mid}{\Gamma}\right\},
\label{fourteen2}
\end{eqnarray}
where $\mid{\varepsilon_f^{*}}\mid$ is the effective position of the
localized level below the chemical potential and the levels width
$\Gamma\propto{V^2\nu{(\epsilon_F)}}$
depends on the hybridization
matrix element,$V$, and the density of states at the Fermi level,
$\nu{(\epsilon_F)}$. The KVC model 
\cite{Allen} connects the first order transition with
strong non-linear dependence of the Kondo scale (\ref{fourteen2})
($\mid{\varepsilon_f^{*}}\mid\ll{\Gamma}$) on the volume through
the volume dependence of the hybridization matrix element
(in Ce change in the unit cell volume  
is large, $\delta{v}/v\sim{20\%}$!). 

Nevertheless, the KVC model seems not to be applicable in case of YbInCu$_4$,
where the volume changes are extremely small 
\cite{Sarrao,Immer}.
On that reason, the FKR model has recently been 
revisited in \cite{Zlatic}. It is interesting, that although
being somewhat sensitive to the choice of the model parameters,
the elliptic shape for the phase transition line 
in the $(B,T)$-plane, observed in \cite{Immer1} 
is preserved in the calculations \cite{Zlatic}. 
This is probably due to the same mechanism 
as above , i.e. due to large differences in the
energy scales for the two phases (it seems however that the
constant $a$ in (\ref{one}) strongly depends on the parameters choice).

As it was discussed in \cite{Dzero}, the universality of the first-order
transition line for YbInCu$_4$ and its alloys in the $B-T$ plane can be 
described in terms of an entropy first-order transition between
the local f-moment phase and another phase with a compensated moment.
It was also suggested in \cite{Dzero} that the mixed-valence transition
is driven by the change in the electronic screening: high-temperature phase
can be described as a band-like semimetal with a small carrier concentration
and accordingly screening is weak, what favors to localization of the 
$f$-electrons. At lower temperatures after a phase transition occurred, 
even the $f$-electrons form a band state, so that a small change in 
occupation numbers would not contradict to an emergence of a large
$f$-like Fermi surface. 

In this paper we would like to address the issue of how 
the phase diagram of YbInCu$_4$ in the $B-T$ plane 
is affected by taking into account the crystal field effects
and, as a consequence, an appearance of anisotropy of a phase
diagram with respect to the direction of an applied field. 
We also would like to analyze the relation, experimentally obtained 
in \cite{Sarrao,Immer}:
\begin{eqnarray}
a\mu_B{B_{c0}} = T_{v0}
\label{one}
\end{eqnarray}
in terms of the crystal field Hamiltonian:
\begin{eqnarray}
\hat{H}&=&\hat{H}_{crystal} + g_J\mu_B{\hat{\bf J}}\cdot{\bf B}.
\label{two}
\end{eqnarray}

When magnetic ion is placed in a cubic environment, the spatial
degeneracy of its angular momentum is removed by the electrostatic
field due to the neighboring charges. For example, the $J=5/2$ 
multiplet for a Ce ion is split into a $\Gamma_7$ doublet 
and $\Gamma_8$ quartet while the $J=7/2$ multiplet of an Yb ion is split 
into a $\Gamma_6$ doublet, $\Gamma_7$ doublet and $\Gamma_8$ quartet. 

For the $J=7/2$ the wave functions for representations $\Gamma_6$, 
$\Gamma_7$ and $\Gamma_8$ are given by ~\cite{Lea}:
\begin{eqnarray}
\Gamma_6 &:& \left\{{\begin{array} 
		{r@{\quad}}
\psi_{1} = \sqrt{\frac{5}{12}}|+\frac{7}{2}\rangle + 
	   \sqrt{\frac{7}{12}}|-\frac{1}{2}\rangle\\
\psi_{2} = \sqrt{\frac{5}{12}}|-\frac{7}{2}\rangle + 
	   \sqrt{\frac{7}{12}}|+\frac{1}{2}\rangle 
           \end{array}} \right\},\label{gamma6}\\
\Gamma_7 &:& \left\{{\begin{array} 
		{r@{\quad}}
\psi_{3} = \frac{\sqrt{3}}{2}|+\frac{5}{2}\rangle - 
	   \frac{1}{2}|-\frac{3}{2}\rangle\\
\psi_{4} = \frac{\sqrt{3}}{2}|-\frac{5}{2}\rangle - 
	   \frac{1}{2}|+\frac{3}{2}\rangle
           \end{array}} \right\},\label{gamma7}\\
\Gamma_8 &:& \left\{{\begin{array} 
		{r@{\quad}}
\psi_{5} = \sqrt{\frac{7}{12}}|+\frac{7}{2}\rangle - 
	   \sqrt{\frac{5}{12}}|-\frac{1}{2}\rangle\\
\psi_{6} = \sqrt{\frac{5}{12}}|-\frac{7}{2}\rangle - 
	   \sqrt{\frac{7}{12}}|+\frac{1}{2}\rangle\\ 
\psi_{7} = \frac{1}{2}|+\frac{5}{2}\rangle + 
	   \frac{\sqrt{3}}{2}|-\frac{3}{2}\rangle\\
\psi_{8} = \frac{1}{2}|-\frac{5}{2}\rangle + 
	   \frac{\sqrt{3}}{2}|+\frac{3}{2}\rangle
           \end{array}} \right\}\label{gamma8}.
\end{eqnarray}

According to \cite{Dzero} the first-order transition line in the 
$(B,T)$ plane is determined by the equation:
\begin{eqnarray}
T\cdot{S(B,T)}&=&const.
\label{entropy}
\end{eqnarray}
The entropy is determined by the Yb$^{3+}$ multiplet structure
only. Taking the crystal splitting effects into account, the entropy is
given by:
\begin{eqnarray}
T\cdot{S(B,T)}&=&-T\ln\left\{\sum\limits_{n=1}^{8}
\exp\left(-\frac{\lambda_n}{T}\right)\right\},
\end{eqnarray}
where $\lambda_n$ are the eigenvalues, obtained by a solution
of an eigenvalue problem for (\ref{two}) on wavefunctions 
(\ref{gamma6}-\ref{gamma8}).

Now we have to find eigenvalues $\lambda_n$. Re-writing the last
term in (\ref{two}) as:
\begin{eqnarray}
g_J\mu_B{\hat{\bf J}}\cdot{\bf B}=\hat{J}_z\beta_z + 
\hat{J}_{+}\beta_{-} + \hat{J}_{-}\beta_{+},
\end{eqnarray} 
where $\hat{J}_{\pm}=\hat{J}_x\pm{i}\hat{J}_y$ and 
$\beta_{\pm}=\frac{g_J\mu_B}{2}\left(B_x\pm{i}B_y\right)$, 
$\beta_z=g_J\mu_B{B_z}$, matrix elements $H_{ij}$ of (\ref{two}) 
are given by:
\end{multicols}
\widetext
\begin{eqnarray}
H_{ij} = \left| \begin{array}{llllllll}

E_6 + \frac{7}{6}\beta_z & \frac{7}{3}\beta_{-} & 0 & 0 & 
\frac{\sqrt{35}}{3}\beta_z & -\frac{7}{3}\beta_{-} & 
\sqrt{\frac{35}{3}}\beta_{+} & 0 \\

\frac{7}{3}\beta_{+} & E_6 - \frac{7}{6}\beta_z & 0 & 0 &
-\frac{\sqrt{35}}{3}\beta_{+} & -\frac{\sqrt{35}}{3}\beta_z & 0 
& \sqrt{\frac{35}{3}}\beta_{-} \\

0 & 0 & E_7 + \frac{3}{2}\beta_z & -3\beta_{+} & 
3\beta_{-} & 0 & \sqrt{3}\beta_z & \sqrt{3}\beta_{+} \\

0 & 0 & -3\beta_{-} & E_7 - \frac{3}{2}\beta_z &
0 & \frac{\sqrt{35}}{2}\beta_{+} & \sqrt{3}\beta_{-} & -\sqrt{3}\beta_z \\

\frac{\sqrt{35}}{3}\beta_z & -\frac{\sqrt{35}}{3}\beta_{-} & 
3\beta_{+} & 0 & E_8 + \frac{11}{6}\beta_z & 
\frac{\sqrt{35}}{3}\beta_{-} & -\frac{2}{\sqrt{3}}\beta_{+} & 0 \\

-\frac{7}{3}\beta_{+} & -\frac{\sqrt{35}}{3}\beta_z & 0 & 
\frac{\sqrt{35}}{2}\beta_{-} & \frac{\sqrt{35}}{3}\beta_{+} & 
E_8 - \frac{11}{6}\beta_z & 0 & -\sqrt{\frac{35}{12}}\beta_{-} \\

\sqrt{\frac{35}{3}}\beta_{-} & 0 & \sqrt{3}\beta_z & \sqrt{3}\beta_{+} &
-\frac{2}{\sqrt{3}}\beta_{-} & 0 & E_8 - \frac{1}{2}\beta_z & 3\beta_{+} \\

0 & \sqrt{\frac{35}{3}}\beta_{+} & \sqrt{3}\beta_{-} & -\sqrt{3}\beta_z & 
0 & -\sqrt{\frac{35}{12}}\beta_{+} & 3\beta_{-} & E_8 + \frac{1}{2}\beta_z 

\end{array} \right|,
\label{matrix}
\end{eqnarray}
\begin{multicols}{2}
\narrowtext
where the matrix elements of $\hat{H}_{crystal}$ are defined as:
\begin{eqnarray}
<\Gamma_6|\hat{H}_{crystal}|\Gamma_6>&=&E_6, 
~<\Gamma_7|\hat{H}_{crystal}|\Gamma_7>=E_7,\nonumber\\
~&&<\Gamma_8|\hat{H}_{crystal}|\Gamma_8>=E_8,
\end{eqnarray}
As it turns out, the secular equation:
\[
det||H_{ij}-\lambda\delta_{ij}||=0,
\]
with Hamiltonian matrix, 
given by (\ref{matrix}) can not be solved exactly in its general
form, but there exist the exact solutions for particular
cases, such as ${\bf B}=(0,0,B_z)$. 
The eigenvalues in that case are:
\begin{eqnarray}
\lambda_{1,2}(\beta_z)&=&\frac{1}{2}\left\{E_6 + E_8 + \frac{24}{7}\beta_z\pm
\right.\nonumber\\&&\left.
\sqrt{\left(E_6-E_8-\frac{16}{21}\beta_z\right)^2 + \frac{8960}{441}\beta_z^2}
\right\},
\label{e1}\\
\lambda_{3,4}(\beta_z)&=&\frac{1}{2}\left\{E_6 + E_8 - \frac{24}{7}\beta_z\pm
\right.\nonumber\\&&\left.
\sqrt{\left(E_6-E_8+\frac{16}{21}\beta_z\right)^2 + \frac{8960}{441}\beta_z^2}
\right\},\label{e2}\\
\lambda_{5,6}(\beta_z)&=&\frac{1}{2}\left\{E_7 + E_8 + \frac{8}{7}\beta_z\pm
\right.\nonumber\\&&\left.
\sqrt{\left(E_7-E_8+\frac{16}{7}\beta_z\right)^2 + \frac{768}{49}\beta_z^2}
\right\},\label{e3}\\
\lambda_{7,8}(\beta_z)&=&\frac{1}{2}\left\{E_7 + E_8 - \frac{8}{7}\beta_z\pm
\right.\nonumber\\&&\left.
\sqrt{\left(E_7-E_8-\frac{16}{7}\beta_z\right)^2 + \frac{768}{49}\beta_z^2}
\right\}.
\label{e4}
\end{eqnarray}
In what follows we present the numerical result, when external field is in
the plane ${\bf B}=(B_x, B_y, 0)$ and analytical solution for the case 
mentioned above.

Now, we can define a constant in Eq. (\ref{entropy}) as follows:
\begin{eqnarray}
&&T\cdot{S(B,T)}=T_{v0}\cdot{S_{0}(0,T_{v0})},
\label{phdiag}\\
&&S_{0}=\ln\left[4 + 2\exp\left({-\frac{\delta{E_{6,8}}}{T_{v0}}}\right) + 
2\exp\left({-\frac{\delta{E_{7,8}}}{T_{v0}}}\right)\right],
\nonumber
\end{eqnarray}
According to \cite{Severing} $\delta{E_{6,8}}=E_6-E_8\simeq{3.2}$meV and 
$\delta{E_{7,8}}=E_7-E_8\simeq{3.8}$meV. 
\begin{figure}[h]
\centerline{\psfig{file=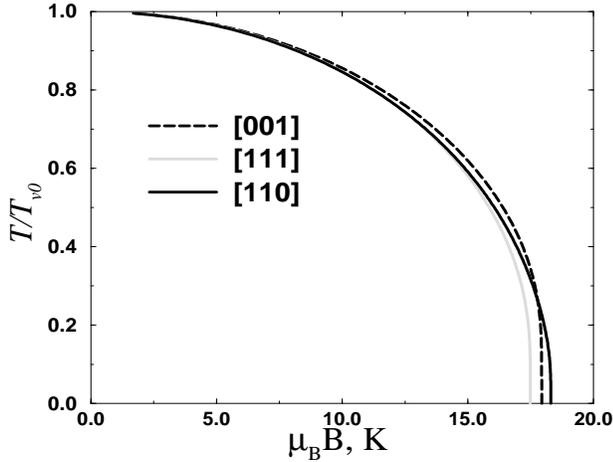,height=7cm,width=9cm,angle=-90}}\
\vspace{0.3cm}
\caption{The phase diagram for the YbInCu$_4$ shows some
anisotropy with respect to the direction of an external magnetic
field: {\bf B} is along one of the main cubic axes (dashed line),
{\bf B} is in the {\it easy plane} (solid black line)
and {\bf B} is along the one of the main cubic diagonals (solid gray line).}
\end{figure}
As we already mentioned, the equation (\ref{phdiag}) defines a phase
diagram in the $B-T$ plane. The results of our calculation are plotted
on Fig. 1. As we can see, there is a strong anisotropy in a phase diagram 
with respect to the direction of an external magnetic field. We also have
calculated magnetization as a function of an external magnetic field for 
a given temperature, Fig. 2. As it turns out, magnetization also
depends on the direction of an applied field.  

In the rest of the paper, we would like to write down an equation 
for the phase boundary, when ${\bf B}=(0,0,B_z)$. 
Let us introduce the following notations:
\[
b=B/B_{v0}, ~\tau=T/T_{v0}, ~\tan(\varphi)=b/\tau 
~(0<\varphi<\frac{\pi}{2}).
\label{param}
\]
Then 
\begin{eqnarray}
\tau&=&\tilde{U}^{-1}(\varphi),\\
b&=&\tilde{U}^{-1}(\varphi)\tan(\varphi),\\
\tilde{U}^{-1}(\varphi)&=&\frac{1}{S(0,T_{v0})}\sum\limits_{n=1}^{8}
\exp\left[-\tilde{\lambda}_n(\varphi)\right],
\end{eqnarray} 
with $\tilde{\lambda}_n(\varphi)$ being the eigenvalues (\ref{e1}-\ref{e4}).
Thus, the equation for the phase transition line is given by:
\[
b^2 + \tau^2=R(\varphi), 
~R(\varphi)=\left[\cos(\varphi)\tilde{U}(\varphi)\right]^{-2}.
\]

As we see on Fig. 3, deviations of $R(\varphi)$ from 1 do not exceed 
10\%. 
Generally speaking, as one may see from Table 1, the $a$ value is 
relatively robust with respect to including new type of interactions
(or anisotropy) in our model. For example, if one takes the 
susceptibility of a lower phase into account, it decreases $a$ by reducing 
the value for the net moment $\mu$ which in our case is less then 
$4\mu_{B}$ and the latter corresponds to the free ion model, but 
on the other hand it increases one by reducing a change in the entropy 
at $T=T_{v0}$. As we see from results of our calculations of 
$B_{c0}$ (Table 1), 
when only crystal fields effect are taken, the change in $a$ is less
then 10\%.
\vspace{-0.5cm}
\begin{figure}[h]
\centerline{\psfig{file=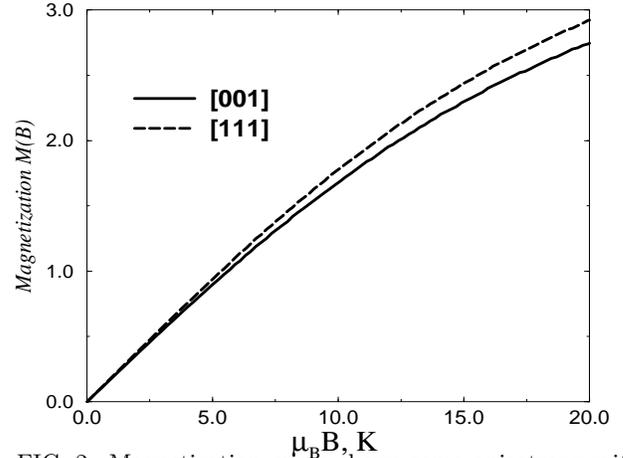,height=7cm,width=9cm,angle=-90}}
\caption{Magnetization curve shows some anisotropy with
respect to the direction of an external magnetic field in 
agreement with our result for the phase diagram where the same
type of anisotropy is found. Calculations were performed for 
$T=0.75T_{v0}$}
\end{figure}
We have to mention that although the crystal field effects do not
change a value of $a$ parameter in (\ref{one}), 
it is not the case if one will try to verify the 
relation (\ref{one}) in terms of exact solution for the Kondo model 
with $J=7/2$ \cite{Schlottmann}. As it turns out, in that case, 
the value for $a$ strongly depends on the 
energy scale, given by $T_K$. This result is in fact similar to the result
obtained by \cite{Zlatic} using the Dynamical Mean Field Theory (DMFT) 
approach based on the FRK model. Thus, experimentally verified 
exsistance of anisotropy in the phase diagram will serve as a 
proof of our initial idea of free Yb ions model.

\vspace{-1cm}
\begin{figure}[h]
\psfig{file=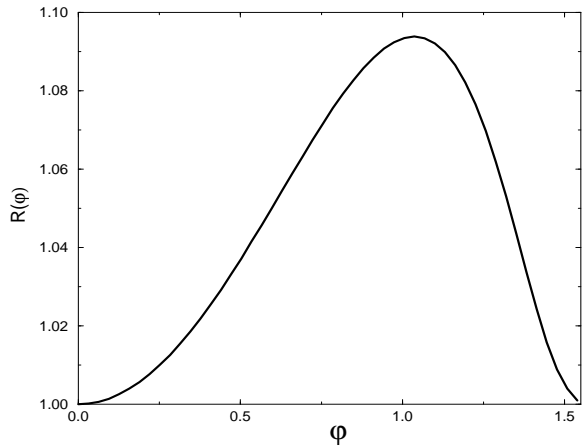,height=7cm,width=9cm,angle=-90}
\caption{Deviations of $R(\varphi)$ from 1 do not exceed 10\%.}
\end{figure}
To summarize, we have shown that the phase diagram for the first 
order valence transition in YbInCu$_4$ in the $B-T$ plane
acquires some anisotropy with respect to the direction of an 
external magnetic field if crystal-field split multiplets are taken 
into consideration. We have also found that within the present framework,
that the anisotropy of the critical field value $B_{c0}$ is of the order
of $\simeq{2}$ Tesla and in principle can be seen experimentally.  
\begin{table}[h]
\caption{Values of $B_{c0}$ and $a$ obtained from the phase diagram Fig. 1 for
different magnetic field orientations.}
\begin{tabular}{|c|c|c|}
Field Orientation & $B_{c0}$, Tesla & a\\
\hline
{[}001{]} & 30.9 & 2.29\\
{[}111{]} & 30.1 & 2.34\\
{[}110{]} & 31.95 & 2.24\\
\end{tabular}
\end{table}

Preliminary experiments have been carried
out in the National High Magnetic Field Laboratory (Tallahassee) and
showed relatively good agreement with theoretical prediction in terms
of the present model \cite{Fisk}.  

I am indebted to A. K. Zvezdin and L. P. Gor'kov for bringing this 
problem to my attention and for very useful comments on this paper.
I also thank P. Schlottmann for providing me with his unpublished 
data regarding the exact numerical solutions for the asymmetric 
Anderson model and S. Nakatsuji for useful discussions. 
This work was supported by the NHMFL through the NSF cooperative 
agreement DMR-9527035 and the State of Florida.

\end{multicols}

\end{document}